\documentclass[sigconf]{acmart}

\usepackage{graphicx}
\usepackage{amsmath}
\usepackage{mathtools,amssymb}
\usepackage{booktabs} % For formal tables
\usepackage{color}
\usepackage{soul}
\usepackage[skip=5pt]{caption}

\usepackage{hyperref}

\setcopyright{rightsretained}
\acmConference[SPIRE'19]{SPIRE}{October 7-9, 2019}{Segovia, Spain}
\acmYear{2019}
\acmDOI{10.1007/978-3-030-32686-9\_4}
\copyrightyear{2019}

\begin{document}

\title{Position Bias Estimation for Unbiased Learning-to-Rank in eCommerce Search}

\author{Grigor Aslanyan}
\affiliation{%
  \institution{eBay Inc.}
  \streetaddress{2025 Hamilton Avenue}
  \city{San Jose}
  \state{CA}
  \postcode{95125}
}
\email{gaslanyan@ebay.com}

\author{Utkarsh Porwal}
\affiliation{%
  \institution{eBay Inc.}
  \streetaddress{2025 Hamilton Avenue}
  \city{San Jose}
  \state{CA}
  \postcode{95125}
}
\email{uporwal@ebay.com}

\begin{abstract}
	The Unbiased Learning-to-Rank framework \cite{Joachims:2017:ULB:3018661.3018699} has been recently proposed as a general approach to systematically remove biases, such as position bias, from learning-to-rank models. The method takes two steps - estimating click propensities and using them to train unbiased models. Most common methods proposed in the literature for estimating propensities involve some degree of intervention in the live search engine. An alternative approach proposed recently uses an Expectation Maximization (EM) algorithm to estimate propensities by using ranking features for estimating relevances \cite{Wang:2018:PBE:3159652.3159732}. In this work we propose a novel method to directly estimate propensities which does not use any intervention in live search or rely on modeling relevance. Rather, we take advantage of the fact that the same query-document pair may naturally change ranks over time. This typically occurs for eCommerce search because of change of popularity of items\footnote{Here an ``item'' is the same as a ``document''. In this work we use the terms ``item'' and ``document'' interchangeably.} over time, existence of time dependent ranking features, or addition or removal of items to the index (an item getting sold or a new item being listed). However, our method is general and can be applied to any search engine for which the rank of the same document may naturally change over time for the same query. We derive a simple likelihood function that depends on propensities only, and by maximizing the likelihood we are able to get estimates of the propensities. We apply this method to eBay search data to estimate click propensities for web and mobile search and compare these with estimates using the EM method \cite{Wang:2018:PBE:3159652.3159732}. We also use simulated data to show that the method gives reliable estimates of the ``true'' simulated propensities. Finally, we train an unbiased learning-to-rank model for eBay search using the estimated propensities and show that it outperforms both baselines - one without position bias correction and one with position bias correction using the EM method.
\end{abstract}

\maketitle

\section{Introduction}

%Introduction should answer the following questions
%1. What is the problem you are trying to solve
%2. Why is this problem important
%3. What has been done to address this issue (some recent work)
%4. What is lacking in what has been done so far
%5. What are you proposing that adds value - Some folks argue this point should be mentioned right away

Modern search engines rely on machine learned methods for ranking the matching results for a given query. Training and evaluation of models for ranking is commonly known as \textbf{Learning-to-Rank (LTR)} \cite{Hang2011ASI}. There are two common approaches for collecting the data for LTR - \textbf{human judgements} and \textbf{implicit user feedback}. For human judgements samples of documents are gathered for a sample of queries and sent to human judges who analyze and label each document. The labels can be as simple as \textit{relevant} vs. \textit{not relevant} or can involve more levels of relevance. This labeled data is then used for training and/or evaluation of LTR models. Collecting human judged data can be expensive and time consuming and often infeasible. On the other hand, data from implicit user feedback, such as clicks, is essentially free and abundant. For that reason it is often the preferred method for collecting data for LTR. A major drawback of this method is that the data can be heavily biased. For example, users can only click on documents that have been shown to them (presentation bias) and are more likely to click on higher ranked documents (position bias). A lot of work in the LTR literature has focused on accounting for and removing these biases. In particular, the recent paper by Joachims et. al. \cite{Joachims:2017:ULB:3018661.3018699} has proposed a framework for systematically removing the biases from user feedback data. Following the title of the paper we will refer to this framework as \textbf{Unbiased Learning-to-Rank}. In particular, the authors have focused on removing the position bias by first estimating the click propensities and then using the inverse propensities as weights in the loss function. They have shown that this method results in an unbiased loss function and hence an unbiased model.

Unbiased Learning-to-Rank is an appealing method for removing the inherent biases. However, to apply it one needs to first get a reliable estimate of click propensities. The method proposed in \cite{Joachims:2017:ULB:3018661.3018699} uses result randomization in the live search engine to estimate propensities. This can negatively impact the quality of the search results, which will in turn result in poor user experience and potential loss of revenue for the company \cite{Wang:2018:PBE:3159652.3159732}. It also adds bookkeeping overhead. Wang et. al. \cite{Wang:2018:PBE:3159652.3159732} have proposed a regression-based Expectation Maximization (EM) method for estimating click propensities which does not require result randomization. However, this method uses the ranking features to estimate relevances and can result in a biased estimate of propensities unless the relevance estimates are very reliable, which is difficult to achieve in practice.

In this paper we propose a novel method for estimating click propensities without any intervention in the live search results page, such as result randomization. We use query-document pairs that appear more than once at different ranks to estimate click propensities. In comparison to the EM-based algorithm in \cite{Wang:2018:PBE:3159652.3159732} our method does not rely on modeling the relevance using ranking features. In fact, we completely eliminate the relevances from the likelihood function and directly estimate the propensities by maximizing a simple likelihood function.

Agarwal et. al. \cite{agarwal2019estimating} have proposed a similar approach for estimating propensities without interventions, which has been done in parallel to our work. The approach developed there relies on having multiple different rankers in the system, such as during A/B tests. They also derive a likelihood function to estimate the propensities, called an \emph{AllPairs} estimator, which depends on terms for all combinations of rank pairs. In comparison to the method in \cite{agarwal2019estimating} our method is more general and does not rely on having multiple rankers in the system. Although requiring multiple rankers is better than intervention it may still have a similar cost. For example, a different ranker could result in a different user experience and extra book keeping overhead. In contrast, our proposed approach leverages the organic ranking variation because of time dependent features and does not result in extra costs. That said, our method can naturally take advantage of having multiple rankers, if available. More importantly, our likelihood function depends on the propensities only, rather than terms for all combinations of pairs. The number of unknown parameters to estimate for our method is linear, rather than quadratic, in the number of ranks, which is a major advantage. Our method can therefore give reliable estimates for much lower ranks using much less data.

We use simulated data to test our method and get good results. We then apply our method on actual data from eBay search logs to estimate click propensities for both web and mobile platforms and compare them with estimates using the EM method \cite{Wang:2018:PBE:3159652.3159732}. Finally, we use our estimated propensities to train an unbiased learning-to-rank model for eBay search and compare it with two baseline models - one which does not correct for position bias and one which uses EM-based estimates for bias correction. Our results show that both unbiased models significantly outperform the ``biased'' baseline on our offline evaluation metrics, with our model also outperforming the EM method \cite{Wang:2018:PBE:3159652.3159732}.

The main novel contributions of this work can be summarized as follows:
\begin{itemize}
	\item We present a new approach for directly estimating click propensities without any interventions in live search. Compared with other approaches in the literature \cite{agarwal2019estimating,Wang:2018:PBE:3159652.3159732}, our approach does not require multiple rankers in the system and large amounts of data for each pair of ranks from different rankers. Moreover, our proposal gives direct estimates of the propensity without having to model relevance. This makes our approach more robust and general.
	\item Under a mild assumption we derive a simple likelihood function that depends on the propensities only. This allows for propensity estimation for much lower ranks. We also prove the validity of the method through simulations.
	\item We estimate propensities up to rank $500$ using our method for a large eCommerce search engine. This is a much lower rank than previous methods in the literature have been able to obtain (around rank $20$). This may not be important for some search engines but is especially important in the eCommerce domain where people typically browse and purchase items from much lower ranks than for web search.
	\item To the best of our knowledge this is the first paper to do a detailed study of the unbiased learning-to-rank approach for eCommerce search.
\end{itemize}

The rest of the paper is organized as follows. In Section \ref{section-related-work} we discuss some of the related work in the literature. In Section \ref{section-method} we introduce our method for estimating click propensities. In Section \ref{sec-ebay} we apply our method to eBay search logs and estimate propensities for web and mobile search, and compare them with EM-based estimates. In Section \ref{section-models} we train and evaluate unbiased learning-to-rank models for eBay search using our estimated propensities as well as the propensities estimated with the EM method \cite{Wang:2018:PBE:3159652.3159732}, and show that the our model outperforms both baselines - one without position bias correction and one with bias correction using estimates from the EM method. We summarize our work in Section \ref{section-summary} and discuss future directions for this research.

We give a brief summary of the Unbiased Learning-to-Rank framework in Appendix \ref{section-unbiased-ltr}. The derivation of our likelihood function is presented in Appendix \ref{section-likelihood-simplification}. In Appendix \ref{sec-ratio} we discuss a simple method for estimating propensities in the case when there is a large amount of data for each pair of ranks of interest. Finally, in Appendix \ref{section-sim} we apply our method to simulated data and show that we are able to obtain reliable estimates of the ``true'' simulated propensities.

\section{Related Work}\label{section-related-work}
Implicit feedback such as clicks are commonly used to train user facing machine learned systems such as ranking or recommender systems. Clicks are preferred over human judged labels as they are available in plenty, are available readily and are  collected in a natural environment. However, such user behavior data can only be collected over the items shown to the users. This injects a presentation bias in the collected data. This affects the machine learned systems as they are trained on user feedback data as positives and negatives. It is not feasible to present many choices to the user and it affects the performance of these systems as we can not get an accurate estimate of positives and negatives for training with feedback available only on selective samples. This situation is aggravated by the fact that the feedback of the user not only depends on the presentation, it also depends on where the item was presented. This is a subclass of the presentation bias called position bias. Joachims et al. \cite{Joachims:2017:ULB:3018661.3018699}  proved that if the collected user behavior data discounts the position bias accurately then the learned system will be the same as the one learned on true relevance signals.

Several approaches have been proposed to de-bias the collected user behavior data. One of the most common approaches is the use of click models. Click models are used to make hypotheses about the user behavior and then the true relevance is estimated by optimizing the likelihood of the collected clicks. There are several types of click models. One such model is a random click model (RCM) \cite{dupret2008user} where it is assumed that every document has the same probability of getting clicked and that probability is the model parameter. In a rank based click through rate model (RCTR) it is assumed that the probability of every document being clicked depends on its rank. Therefore, the total number of model parameters is the total number of ranks in the ranking system. Another model is the document based CTR model (DCTR) \cite{craswell2008experimental} where the click through rates are estimated for each query-document pair. In this model the total number of model parameters is the total number of query-document pairs. This model is prone to overfitting as the number of parameters grows with the training data size. Most commonly used click models are the position based model (PBM) \cite{craswell2008experimental,Joachims2005} and the cascade model (CM) \cite{craswell2008experimental}. In PBM the hypothesis is that a document is only clicked if it is observed and the user found it attractive or relevant. In CM the hypothesis is that the user sequentially scans the whole document top to bottom and clicks when the document is found to be relevant. In this model the top document is always observed and consecutive documents are only observed if the previous ones were observed and were not deemed relevant. In our proposed method we make a similar hypothesis such as the position based method where the observation probability depends on the rank and the probability of relevance only depends on the query-document pair. However, our approach is to learn the click propensities instead of learning the true relevance by optimizing the likelihood of the collected clicks. More advanced click models, such as the user browsing model (UBM) \cite{dupret2008user}, the dependent click model (DCM) \cite{guo2009efficient}, the click chain model (CCM) \cite{guo2009click}, and the dynamic Bayesian network model (DBN) \cite{chapelle2009dynamic} are also proposed. Chuklin et al. \cite{chuklin2015click} provides a comprehensive overview of click models.

Click models are trained on the collected user behavior data. Interleaving is another option that is deployed at the time of data collection. In interleaving different rank lists can be interleaved together and presented to the user. By comparing the clicks on the swapped results one can learn the unbiased user preference. Different methods for interleaving have been proposed. In the balanced interleave method \cite{Joachims03evaluatingretrieval} a new interleaved ranked list is generated for every query. The document constraint method \cite{he2009evaluation} accounts for the relation between documents. Hofmann et al. \cite{hofmann2011probabilistic} proposed a probabilistic interleaving method that addressed some of the drawbacks of the balanced interleave method and the document constraint method. One limitation of the interleaving method is that often the experimentation platform in eCommerce companies is not tied to just search. It supports A/B testing for all teams, such as checkout and advertisements. Therefore, the interleaving ranked list may not be supported as it is pertinent only for search ranking.

A more recent approach to address presentation bias is the unbiased learning-to-rank approach. In this click propensities are estimated and then the inverse propensities are used as weights in the loss function. Click propensities are estimated by presenting the same items at different ranks to account for click biases without explicitly estimating the query-document relevance. Click propensity estimation can either be done randomly or in a more principled manner. Radlinski et al. \cite{radlinski2006minimally} presented the FairPairs algorithm that randomly flips pairs of results in the ranking presented to the user. They called it randomization with minimal invasion. Carterette et al. \cite{Carterette2018} also presented a minimally invasive algorithm for offline evaluation. Joachims et al. \cite{Joachims:2017:ULB:3018661.3018699} proposed randomized intervention to estimate the propensity model. Radlinski et al. \cite{radlinski2008learning}, on the other hand, proposed alteration in ranking in a more informed manner using Multi-Armed Bandits. The main drawback of randomization for propensity estimation is that it can cause bad user experience, book keeping overhead, and a potential loss in revenue. Wang et al. \cite{Wang:2018:PBE:3159652.3159732} proposed a method to estimate propensities without randomization using the EM algorithm. In most of the existing methods, propensity estimation is done first. Once the propensities are learned, an unbiased ranker is trained using the learned propensities. Recently Ai et al. \cite{ai2018unbiased} proposed a dual learning algorithm that learns an unbiased ranker and the propensities together.

\section{Propensity Estimation Method}\label{section-method}

The method proposed by Joachims et. al. \cite{Joachims:2017:ULB:3018661.3018699} for estimating click propensities is running an experimental intervention in the live search engine, where the documents at two selected ranks are swapped. By comparing the click through rates at these ranks before and after swapping one can easily estimate the ratios of propensities at these ranks (one only needs the ratio of propensities for removing the position bias \cite{Joachims:2017:ULB:3018661.3018699}). Here we propose a novel methodology for estimating click propensities without any intervention. For some search engines, especially in eCommerce, the same query-document pair may naturally appear more than once at different ranks. Using the click data on such documents we can accurately estimate click propensities. It is not required that the same query-document pair should appear at different ranks a large number of times.

We model clicks by the following simple model (also used in \cite{Joachims:2017:ULB:3018661.3018699}) - \emph{The probability of a click on a given document is the product of the probability of observing the document and the probability of clicking on the document for the given query assuming that it has been observed.} We assume that the probability of observing a document depends only on its rank and the probability of clicking on the document for a given query if it is observed depends only on the query and the document. Mathematically:
\begin{align}\label{prob-click}
\begin{split}
	p(c=1|q,y)&=p(o=1|q,y)p(c=1|q,y,o=1)\\
		  &=p(o=1|rank(y))p(c=1|q,y,o=1)\\
		  &=p_{rank(y)}p(c=1|q,y,o=1)
\end{split}
\end{align}
where $q$ denotes a query, $y$ denotes a document, $c$ denotes a click ($0$ or $1$), $o$ denotes observation ($0$ or $1$), and $p_i$ denotes the propensity at rank $i$.

% In this paper we will focus on position bias. We will assume that the probability of observing a document depends only on the original rank of the document (i.e. the rank at which the document was presented to the user at the time of data collection). We denote the propensity at rank $i$ by $p_i$ for $i\in[1,R_{\max}]$, where $R_{\max}$ is the maximum rank used in the dataset. If these propensities are known then the Unbiased learning-to-rank method described above can be used to remove the position bias from the LTR model.

Let us assume that our data $D$ consists of $N$ query-document pairs $x_j$ for $j\in[1,N]$. For a query-document pair $x_j$ we will denote the probability of clicking on the document after observing it by $z_j$. For each query-document pair $x_j$ we have a set of ranks $r_{jk}$ where the document has appeared for the query, and clicks $c_{jk}$ denoting if the document was clicked or not ($1$ or $0$) when it appeared at rank $r_{jk}$, for $k\in[1,m_j]$. Here we assume that the query-document pair $x_j$ has appeared $m_j$ separate times. For now we do not assume that $m_j$ must be greater than 1 - it can be any positive integer.

The probability of a click for query-document pair $x_j$ where the document appeared at rank $r_{jk}$ is, according to (\ref{prob-click}) $p(c=1)=p_{r_{jk}}z_j$. It follows that
 $p(c=0)=1-p_{r_{jk}}z_j$. We can now introduce the following likelihood function:
\begin{equation}\label{likelihood}
	\mathcal{L}(p_i,z_j|D)=\prod_{j=1}^N\prod_{k=1}^{m_j}\left[c_{jk}p_{r_{jk}}z_j+(1-c_{jk})(1-p_{r_{jk}}z_j)\right]\,.
\end{equation}
Here the parameters are the propensities $p_i$ and the ``relevances'' $z_j$ (relevance here means probability of clicking for a given query-document pair assuming that the document has been observed). Theoretically, the parameters can be estimated by maximizing the likelihood function above. However, this can be challenging due to the large number of parameters $z_j$. In fact, we are not even interested in estimating the $z_j$ - we only need to estimate the propensities $p_i$, and the $z_j$ are nuisance parameters.

The likelihood function above can be simplified under mild and generally applicable assumptions. Firstly, only query-document pairs that appeared at multiple different ranks and got at least one click are of interest. This is because we need to compare click activities for the same query-document pair at different ranks to be able to gain some useful information about propensities with the same ``relevance''. Secondly, we make the assumption that overall click probabilities are not large (i.e. not close to $1$). We discuss this assumption in detail in \ref{section-likelihood-simplification}. As we will see in Section \ref{sec-ebay} this is a reasonable assumption for eBay search. This assumption is generally valid for lower ranks (below the top few), and in Appendix \ref{section-likelihood-simplification} we discuss how to make small modifications to the data in case the assumption is violated for topmost ranks. We also discuss alternative approaches for estimating the click propensities for cases when the our assumption might not work very well (our methodology of simulations in Appendix \ref{section-sim} can be used to verify the validity of the assumption).

The likelihood can then be simplified to take the following form:
\begin{equation}\label{log-like}
	\log\mathcal{L}(p_i|D)=\sum_{j=1}^N\left(\log(p_{r_{jl_j}})-\log\sum_{k=1}^{m_j}p_{r_{jk}}\right)\,.
\end{equation}

The detailed derivation is presented in Appendix \ref{section-likelihood-simplification}. Note that the simplified likelihood function (\ref{log-like}) only depends on the propensities, which is one of the most important contribution of this work. By maximizing the likelihood function above we can get an estimate of the propensities. Because the likelihood function depends on the propensities only we can estimate the propensities up to much lower ranks than previously done in the literature without having to rely on a large amount of data.

\section{Click Propensities for eBay Search}\label{sec-ebay}

\begin{figure}
	\includegraphics[width=8cm]{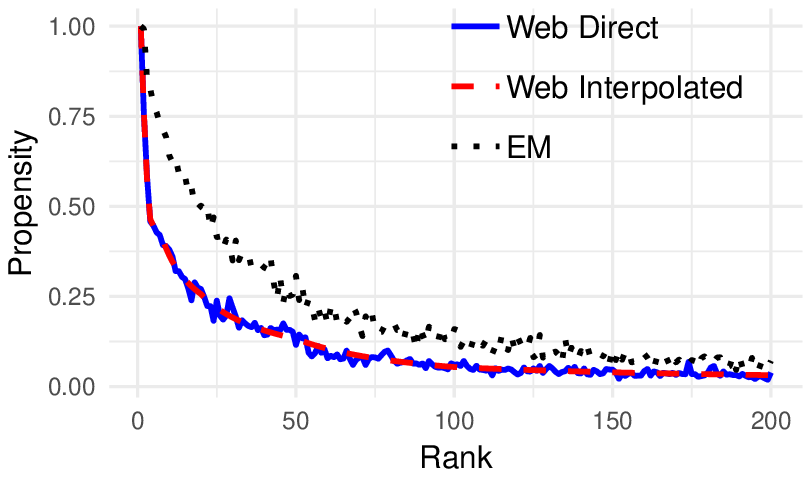}
	\includegraphics[width=8cm]{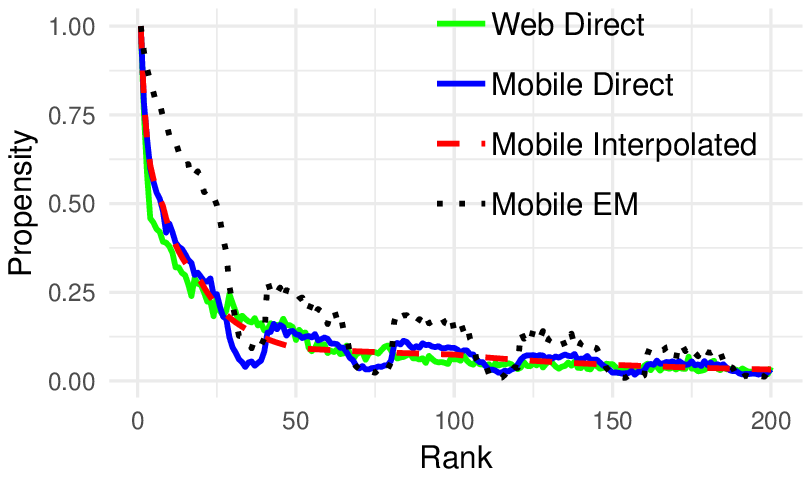}
\setlength{\belowcaptionskip}{-12pt}
\caption{Click propensity estimated for eBay search for web data (top) and mobile data (bottom). The solid blue line is the direct estimation of propensities for each rank, the red dashed line is the estimation using interpolation, and the black dotted curve is the estimation using the EM method. For comparison, on the right side we also plot the propensities for web data using interpolation in solid green, which is the same as the red dashed line from the top plot.}\label{fig-prop-data}
\end{figure}

In this section we apply the method developed above on eBay search data to estimate propensities. For comparison, we also estimate the propensities using the EM method \cite{Wang:2018:PBE:3159652.3159732}.

We collected a small sample (0.2\%) of queries for four months of eBay search traffic. For each query we keep the top 500 items (as mentioned before, we use the terms ``item'' and ``document'' interchangeably). There are multiple sort types on eBay (such as Best Match, Price Low to High, Time Ending Soonest) and click propensities may differ for different sort types. In this paper we present our results on Best Match sort, and hence we keep only queries for that sort type. Furthermore, there are multiple different platforms for search (such as a web browser or a mobile app) which can have different propensities. We separate our dataset into two platforms - web and mobile, and estimate click propensities for each platform separately. For web queries we estimate the propensities for list view with 50 items per page (the most common option).

Next, we identify same query-document pairs and find cases where the document appeared at multiple different ranks. We apply certain filters to ensure that the ``relevance'' of the document has not changed for the query between multiple appearances, and different click probabilities are only due to different ranks. Namely, we check that the price of the item has not changed and exclude auction items (since their relevance depends strongly on the current bid and the amount of time left). We also keep the same query-document pairs from the same day only to make sure that seasonality effects do not affect the popularity of the item. For the query side we identify two queries to be the same if they have the same keywords, as well as the same category and aspect (such as color, size) constraints. We then keep only those query-document pairs that appeared at two different ranks and got one click in one rank and no click in the other.\footnote{Note that keeping only query-document pairs that appeared at two ranks exactly is in no way a requirement of our method. The method is general and can be used for query-document pairs that appeared more than twice.  This is just intended to simplify our analysis without a significant loss in data, since it is rare for the same query-document pair to appear at more than two ranks.} We have also verified our assumption of not very large click probabilities for our dataset. Note that the validity of the assumption is also verified through simulations in Appendix \ref{section-sim} where the simulated data has similar click through rates to the actual eBay data.

We first estimate propensities for web queries. Our dataset consists of about 40,000 query-item pairs, each of which appeared at two different ranks and received a click at one of the ranks. We use two methods for estimating propensities - direct and interpolation. In the direct method we treat the propensity at each rank as a separate parameter. We therefore get 500 different parameters to estimate. In the interpolation method we fix a few different ranks and use the propensities at those ranks as our parameters to estimate. The propensities for all the other ranks are computed as a linear interpolation in the log-log space, i.e. we approximate the log of the propensity as a linear function of the log of the rank. This results in the propensity being a power law of the rank. For the interpolation method our fixed ranks are 1, 2, 4, 8, 20, 50, 100, 200, 300, and 500. We choose a denser grid for higher ranks since there is more data and less noise for higher ranks, and the propensities can be estimated more accurately.

Our resulting propensity for web search is shown in Fig.~\ref{fig-prop-data} (top). The solid blue line shows the propensities estimated through the direct method, and the red dashed curve shows the propensities estimated through interpolation. Even though we estimate propensities up to rank 500, we plot them only up to rank 200 so that the higher ranks can be seen more clearly. The red dashed curve passes smoothly through the blue solid curve, which is reassuring. Note that the red dashed curve is not a fit to the blue one. The two are estimated directly from the data. For the blue curve the parameters are all of the propensities at each rank, whereas for the red dashed curve we only parametrize the propensities at select ranks and interpolate in between. We then maximize the likelihood for each case to estimate the parameters. The fact that the red dashed line appears to be a smooth fit to the solid blue shows that the interpolation method is useful in obtaining a smooth and less noisy propensity curve which is still very close to the direct estimation.

The propensities estimated from eBay mobile search data are shown in Fig.~\ref{fig-prop-data} (bottom). As in the top plot (web data), the blue solid curve shows direct estimation, and the red dashed curve is estimation using interpolation. For comparison, we plot the propensities from web using interpolation in solid green. The blue solid curve shows a certain periodicity - the propensities seem to drop sharply near rank 25, then go back up at rank 40, drop again around rank 65, then back at rank 80, and so on. In fact, this reflects the way results are loaded in mobile search - 40 at a time. The blue curve seems to indicate that users observe the results at higher ranks with the usual decrease in interest, then they tend to scroll faster to the bottom skipping the results towards the bottom, then as the new batch is loaded they regain interest. The red dashed curve matches the blue one reasonable well, but it fails to capture the periodic dips. This is due to our choice of knots for the linear spline. One can use the blue curve to choose new locations of the knots to be able to get a better interpolation for the propensities. The green solid curve matches fairly well with the blue one except for the dips. This means that the propensities for web and mobile are very similar, except for the periodic dips for mobile. The web results are shown 50 items per page, but we have not found any periodic dips for web search. Perhaps this indicates that for web search users do not tend to scroll quickly towards the end of the page and then regain interest as a new page is loaded. The smooth decline in propensities indicates that for web search users steadily lose interest as they scroll down, but the number of items per page does not affect their behavior.

We have also estimated propensities using the regression-based EM method by Wang. et. al. \cite{Wang:2018:PBE:3159652.3159732}. The results are plotted with black dotted lines in Figure \ref{fig-prop-data}. The two methods are very different and use different kinds of data so it is hard to have a fair comparison. However, we have used datasets of similar sizes with similar numbers of queries to make the comparison as fair as possible. For the regression method we have used gradient boosted decision trees \cite{friedman2001} using our top 25 ranking features. The estimates obtained with the EM method are in general higher than the estimates using our method. We have obtained similar periodicity patterns for mobile data from both methods which is reassuring. We do not have the ground truth for comparison since we have not performed any randomization experiments. However, our simulations in the next Section show that our method's predictions are close to the ground truth. We have also used these estimates in Section \ref{section-models} to train unbiased learning-to-rank models and have obtained better offline metrics using our estimates compared to the EM-based estimates.

\section{Unbiased Learning-to-Rank Models}\label{section-models}

In this section we study the improvement in ranking models by using the estimated click propensities for eBay search data. Previous studies have consistently shown that unbiased learning-to-rank models significantly improve ranking metrics compared to their biased counterparts. Specifically, Joachims et. al. \cite{Joachims:2017:ULB:3018661.3018699} have shown that an unbiased learning-to-rank model significantly improves the average rank of relevant results for simulated data. Furthermore, they have performed an online interleaving experiment on a live search engine for scientific articles, which resulted in a significant improvement for the unbiased model. Wang et. al. \cite{Wang:2018:PBE:3159652.3159732} have shown an improvement in MRR (Mean Reciprocal Rank) for the unbiased learning-to-rank models for personal search.

We train ranking models to check if unbiased ranking models show improvements over their biased counterparts and to compare our method of propensity estimation to the EM method. For our training data we collect a sample of about 40,000 queries which have received at least one click. The sample is collected from four days of search logs. We train listwise ranking models using the LambdaMART algorithm \cite{from-ranknet-to-lambdarank-to-lambdamart-an-overview}. We use the DCG metric (\ref{dcg}) as our loss function. We define $\mathrm{rel}_{ij}$ to be $1$ if document $j$ was clicked, and $0$ otherwise. We train three models - one without position bias correction (\emph{baseline biased}), one with position bias correction using propensity estimates from the EM method (\emph{baseline EM}), and finally a model with position bias correction using propensity estimates from our method (\emph{proposed method}). All models use (\ref{dcg}) as a loss function, with \emph{baseline biased} using no position bias correction and the other models using inverse-propensity weighted relevances as in (\ref{loss-unbiased}). We use the propensities estimated for eBay web search as shown in Fig.~\ref{fig-prop-data} (top) - red dashed curve for \emph{proposed method} and black dotted curve for \emph{baseline EM}. Our training and test data are also from web search (i.e. browser) only. We use $25$ features for all models, selected from our top ranking features. We use the same hyperparameters for all the models: the number of trees is $100$ and the shrinkage is $0.1$ (we have fixed the number of trees and tuned the shrinkage for the baseline model, which is then applied to all models).\footnote{Note that these ranking models are significantly different from the eBay production ranker, the details of which are proprietary.}

Our test data contains a sample of about 10,000 queries from four days of eBay search logs. Since the test data also has the same position bias as the training data we cannot rely on standard ranking metrics such as DCG, NDCG (Normalized Discounted Cumulative Gain), or MRR (Mean Reciprocal Rank). Another option would be to use inverse-propensity-weighted versions of these metrics to remove the presentation bias. However, the true propensities are unknown to us and we obviously cannot use estimated propensities for evaluation since part of the evaluation is checking if our estimate of propensities is a good one. For that reason we choose a different approach for evaluation. Namely, we fix the rank of items in the test data, i.e. we select items from different queries that appeared at a given fixed rank. By selecting the items from a fixed rank in the evaluation set we effectively eliminate position bias since all of the items will be affected by position bias the same way (the observation probability is the same for all the items since the rank is the same). Then we compare the two ranking models as classifiers for those items, which means that we evaluate how well the models can distinguish items that were clicked from ones that were not. We use AUC (Area Under the Receiver Operating Characteristic Curve) as our evaluation metric.

\begin{table}[t]
	\begin{center}
		\caption{AUC improvement of the proposed method compared to two baselines - \emph{baseline biased} and \emph{baseline EM} \cite{Wang:2018:PBE:3159652.3159732}. The validation set contains documents from a fixed rank, shown in the first column. The next two columns show the improvements in AUC. Error bars are obtained using 1,000 bootstrap samples of the test data - we show the mean and standard deviation of the improvement over the bootstrap samples.}
		\label{eval-table}
		\begin{tabular}{|c|c|c|c|c|c|}
			\hline
			Rank & Imp. over baseline biased & Imp. over baseline EM \\
			\hline
			$1$  & $3.4\pm1.0\%$ & $1.0\pm0.4\%$ \\
			$2$  & $2.4\pm1.1\%$ & $0.6\pm0.4\%$ \\
			$4$  & $4.2\pm1.2\%$ & $0.7\pm0.4\%$ \\
			$8$  & $3.3\pm1.3\%$ & $1.2\pm0.5\%$ \\
			$16$ & $6.8\pm1.7\%$ & $1.1\pm0.6\%$ \\
			$32$ & $0.8\pm1.8\%$ & $0.8\pm0.7\%$ \\
			\hline
		\end{tabular}
	\end{center}
	\vspace{-6mm}
\end{table}

The results are presented in Table~\ref{eval-table}, where we show results for fixed ranks $1$, $2$, $4$, $8$, $16$, and $32$. To estimate statistical significance of the improvements we have performed 1,000 bootstrap samples of the test data and computed the improvements on these samples. In Table~\ref{eval-table} we show the mean and standard deviation on the bootstrap samples (the distribution of the results on the bootstrap samples is close to Gaussian, as expected, so the mean and standard deviation are enough to describe the full distribution). As we can see, for all ranks the \emph{proposed method} outperforms both baselines. Both unbiased models significantly outperform \emph{baseline biased}. However, our \emph{proposed method} outperforms \emph{baseline EM} as well. The improvements are statistically significant for all ranks, except for rank $32$, where the improvements are not as large. For ranks below $32$ the improvements become minor.

\section{Summary and Future Work}\label{section-summary}

In this work we have introduced a new method for estimating click propensities for eCommerce search without randomizing the results during live search. Our method uses query-document pairs that appear more than once and at different ranks. Although we have used eCommerce search as our main example, the method is general and can be applied to any search engine for which ranking naturally changes over time. The clear advantage of our method over result randomization is that it does not affect live search results, which can have a negative impact on the engine as has been shown in the literature \cite{Wang:2018:PBE:3159652.3159732}. We have compared our method to the EM (Expectation Maximization) based method proposed in \cite{Wang:2018:PBE:3159652.3159732} and have shown that our proposed method outperforms the EM based method for eBay data. There is another approach proposed in parallel to our work \cite{agarwal2019estimating} for direct estimation of propensities. However, our method has a few clear advantages, such as not relying on multiple rankers in the system and not requiring a large amount of data for each pair of ranks. This has allowed us to estimate propensities up to ranks that are much lower than previously computed in the literature. Our proposed approach is robust and we believe that it will find widespread use for unbiased learning-to-rank modeling, especially in the eCommerce domain.

We have used simulated data to show that our method can give accurate estimates of the true propensities. We have applied our method to eBay search results to separately estimate propensities for web and mobile search. We have also trained ranking models and compared the performance of the unbiased model using the estimated propensities to two baselines - one without bias correction and one that corrects position bias using estimates from the EM method. Using a validation dataset of documents from a fixed rank we have shown that our unbiased model outperforms both baselines in terms of the AUC metric.

The focus of this work is propensity estimation from query-document pairs that appear at multiple different ranks. Importantly, we have addressed the case when the same query-document pair appears only a few number of times at different ranks (can be as few as twice). This method can be generalized to use query-document pairs that appeared at a single rank only by incorporating appropriate priors and using Gibbs sampling to estimate the posterior distribution for propensities. We plan to study this approach in a future work. We are also planning to estimate and compare propensities for different classes of queries (such as queries for electronics versus fashion categories) and user demographics, as well as different sort types, such as sort by price.

\appendix

\section{Unbiased Learning-to-Rank}\label{section-unbiased-ltr}

In this Appendix we give a summary of the unbiased learning to rank methodology from \cite{Joachims:2017:ULB:3018661.3018699}. Note that our notation differs from that of \cite{Joachims:2017:ULB:3018661.3018699}.

First, let us assume that we have unbiased data. The data consists of a sample $\mathbf{Q}$ of i.i.d. (independent and identically distributed) queries. Each query $q_i\in\mathbf{Q}$ comes with a set of documents $\mathbf{Y}_i$ with their known relevances. We denote the documents in $\mathbf{Y}_i$ by $y_{ij}$ and their corresponding relevances by $r_{ij}$. We would like to train a LTR model which is a function $F(q, y)$ that computes a score for a given query $q$ and document $y$. These scores can then be used to rank the documents (higher scored documents will be ranked higher). The model is trained to minimize a \emph{loss function} for the training data. The loss function will have smaller values for ranking functions which give higher ranks to relevant documents and lower ranks to irrelevant documents in the training data. There are multiple approaches to choosing a model $F$ and a loss function.

There are three main classes of LTR models - \textbf{pointwise}, \textbf{pairwise}, and \textbf{listwise} \cite{Hang2011ASI,Cao:2007:LRP:1273496.1273513}. Pointwise LTR models use document level loss functions, regardless of the query. Pairwise models, on the other hand, use pairs of documents with desired orders (e.g. pairs containing one relevant and one irrelevant document). The model tries to produce ranking scores that minimize the number of out-of-order pairs. Finally, listwise models take the whole query into account and will optimize for the ordering of the whole list. 

We assume that the loss function takes the form:
\begin{equation}\label{loss-general}
	\Delta(F) = \sum_{y_{ij}}\delta(y_{ij}|F)
\end{equation}
where the sum is taken over all documents $y_{ij}$ in the training set, and $\delta$ denotes a document level loss function. A simple example of a quadratic loss function would be $\delta_\mathrm{quad}(y_{ij}|F)=(r_{ij}-F(y_{ij}))^2$, where $r_{ij}$ is the relevance of document $y_{ij}$. Another popular choice is the cross-entropy loss function if the relevances are binary ($0$ or $1$): $\delta_{CE}(y_{ij}|F)=-r_{ij}\log F(y_{ij})-(1-r_{ij})\log(1-F(y_{ij}))$.

Note that the form (\ref{loss-general}) does not restrict the loss function to be pointwise - it can also be pairwise or listwise. In fact, it has been shown in the literature that the unbiased learning-to-rank framework works better for pairwise/listwise models than pointwise ones \cite{Joachims:2017:ULB:3018661.3018699,Wang:2018:PBE:3159652.3159732}. We will discuss the reasons behind this later in this Appendix. A popular example of a listwise loss function (and one we will use later for our models) is the DCG (Discounted Cumulative Gain):
\begin{equation}\label{dcg}
	\mathrm{DCG}=\sum_{i,j}\frac{\mathrm{rel}_{ij}}{\log_2(j+1)}
\end{equation}
where $i$ is the index of the query, $j$ is the rank of a given document and $\mathrm{rel}_{ij}$ is its relevance, and the sum is taken over all queries and all documents for each query.

If we had data for a fair sample of queries and a fair sample of documents for these queries then minimizing the loss function (\ref{loss-general}) above would result in an unbiased model (this is known in the literature as Empirical Risk Minimization \cite{Vapnik1998}). However, the data is often biased. For example, if we use click logs to determine relevances then we will only have data for documents that the users have actually seen. Using only that data for training will introduce a bias since some documents are more likely to be seen by a user than others. For example, documents that were ranked highly for a given query are more likely to be seen and receive clicks than documents ranked at lower positions (position bias). If we only use the data for documents for which the relevances have been revealed to us (e.g. the user has seen the document and decided to click or not click) we will end up with a biased model. On the other hand, we have no way of including the data for which the relevances have not been revealed. The Unbiased learning-to-rank methodology \cite{Joachims:2017:ULB:3018661.3018699} introduces a modification to the loss function (\ref{loss-general}) such that it becomes an unbiased estimator for the true loss even if the data is biased. The requirement is that we should know the probabilities of observing the relevances for all of the documents in the data (in the context of this paper observing the relevance means that the user has examined the document and decided to click on it or not). In other words, for each document $y_{ij}$ in the training data we know the probability $p(y_{ij})$ of the relevance of that document being observed. This probability is commonly referred to as the \textbf{propensity}. If the propensities are known then an unbiased estimator of the loss function (\ref{loss-general}) is
\begin{equation}\label{loss-unbiased}
	\hat{\Delta}(F) =  \sum_{y_{ij}}\frac{o(y_{ij})\delta(y_{ij}|F)}{p(y_{ij})}=\sum_{y_{ij}:o(y_{ij})=1}\frac{\delta(y_{ij}|F)}{p(y_{ij})}\,.
\end{equation}
Here $o(y_{ij})$ denotes if the document $y_{ij}$ has been \emph{observed} ($1$ if it has been observed, $0$ otherwise). The document being observed is equivalent to the relevance being revealed to us. The equation above only includes data that has been observed, so it can be used in practice. To show that (\ref{loss-unbiased}) is an unbiased estimator of (\ref{loss-general}) we compute the expected value of (\ref{loss-unbiased}):
\begin{align}\label{unbiased-proof}
	\begin{split}
	\mathbb{E}_{o(y_{ij})}[\hat{\Delta}(F)]&=\sum_{y_{ij}}\mathbb{E}_{o(y_{ij})}\left[\frac{o(y_{ij})\delta(y_{ij}|F)}{p(y_{ij})}\right]
					       =\sum_{y_{ij}}\frac{p(y_{ij})\delta(y_{ij}|F)}{p(y_{ij})}\\
					       &=\sum_{y_{ij}}\delta(y_{ij}|F)
					       =\Delta(F)\,.
	\end{split}
\end{align}

The unbiased loss function (\ref{loss-unbiased}) can be used if we know all of the observed documents, as well as their propensities. However, in practice it might be hard to know all of the observed documents. We know for sure that clicked documents have been observed, but we may not have information about documents that have not been clicked. For this reason it is more desirable to have a loss function that includes only ``relevant'' documents, such as clicked ones. This is why pairwise/listwise models work better than pointwise ones for unbiased learning-to-rank \cite{Joachims:2017:ULB:3018661.3018699,Wang:2018:PBE:3159652.3159732}. For example, the DCG loss function (\ref{dcg}) only includes relevant documents, assuming $\mathrm{rel}_{ij}=0$ for irrelevant ones.

\section{Likelihood Function Simplification}\label{section-likelihood-simplification}

There are multiple approaches that one can take to estimate the propensities depending on the data itself. Let us first consider the query-document pairs that appeared only at one rank. The parameters $p_i$ and $z_j$ appear only as a product of each other in the likelihood function (\ref{likelihood}). These query-document pairs could be helpful in estimating the product of the propensity at the rank that they appeared at and the relevance $z_j$ but not each one individually. With $z_j$ unknown, this would not help to estimate the propensity. We should mention that in the presence of a reliable prior for $z_j$ and/or $p_i$ the likelihood function above can be used even for those query-document pairs that appeared only at one rank. In this case it would be more useful to take a Bayesian approach and estimate the posterior distribution for the propensities, for example using Gibbs sampling \cite{10.2307/2685208}.

From now on we will assume that the query-document pairs appear at least at two different ranks. Another extreme is the case when each query-document pair appears a large number of times at different ranks. This will mean that we will get a large number of query-document pairs at each rank. This case is explored below in Appendix \ref{sec-ratio} where we develop a simple and effective method to estimate propensity ratio without the need to maximize the likelihood.

Let us now consider the case when the data consists of a large number of query-document pairs that appeared a few times (can be as few as twice) at different ranks, but the query-document pairs do not appear a large enough number of times to be able to use the method of Appendix \ref{sec-ratio}. In this case we will actually need to maximize the likelihood above and somehow eliminate the nuisance parameters $z_j$ to get estimates for the $p_i$. We will focus the rest of this work on this case. Also, the data we have collected from eBay search logs falls in this category, as discussed in Section \ref{sec-ebay}.

If a query-document pair appeared only a few times there is a good chance that it did not receive any clicks. These query-document pairs will not help in estimating the propensities by likelihood maximization because of the unknown parameter $z_j$. Specifically, for such query-document pairs we will have the terms $\prod_{k=1}^{m_j}(1-p_{r_{jk}}z_j)$. If we use the maximum likelihood approach for estimating the parameters then the maximum will be reached by $z_j=0$ for which the terms above will be $1$. So the query-document pairs without any clicks will not change the maximum likelihood estimate of the propensities. For that reason we will only keep query-document pairs that received at least one click. However, we cannot simply drop the terms from the likelihood function for query-document pairs that did not receive any clicks. Doing so would bias the data towards query-document pairs with a higher likelihood of click. Instead, we will replace the likelihood function above by a conditional probability. Specifically, the likelihood function (\ref{likelihood}) computes the probability of the click data $\{c_{jk}\}$ obtained for that query-document pair. We need to replace that probability by a conditional probability - the probability of the click data $\{c_{jk}\}$ under the condition that there was at least one click received: $\sum_kc_{jk}>0$. The likelihood function for the query-document pair $x_j$ will take the form:
\begin{align}\label{likelihood_j}
\begin{split}
	\mathcal{L}_j(p_i,z_j|D_j)&=P\left(D_j|\sum_k c_{jk}>0\right)\\
				  &=\frac{P(D_j\cap\sum_k c_{jk}>0)}{P(\sum_k c_{jk}>0)}=\frac{P(D_j)}{P(\sum_k c_{jk}>0)}\\
				  &=\frac{\prod_{k=1}^{m_j}\left[c_{jk}p_{r_{jk}}z_j+(1-c_{jk})(1-p_{r_{jk}}z_j)\right]}{1-\prod_{k=1}^{m_j}(1-p_{r_{jk}}z_j)}\,.
\end{split}
\end{align}
Here $\mathcal{L}_j$ denotes the likelihood function for the query-document pair $x_j$, $D_j=\{c_{jk}\}$ denotes the click data for query-document pair $j$, and $P$ denotes probability. $\sum_k c_{jk} > 0$ simply means that there was at least one click. In the first line above we have replaced the probability of data $D_j$ by a conditional probability. The second line uses the formula for conditional probability. The probability of $D_j$ and at least one click just equals to probability of $D_j$ since we are only keeping query-document pairs that received at least one click. This is how the second equality of the second line is derived. Finally, in the last line we have explicitly written out $P(D_j)$ in the numerator as in (\ref{likelihood}) and the probability of at least one click in the denominator (the probability of no click is $\prod_{k=1}^{m_j}(1-p_{r_{jk}}z_j)$ so the probability of at least one click is $1$ minus that).

The full likelihood is then the product of $\mathcal{L}_j$ for all query-document pairs:
\begin{equation}\label{likelihood-1}
	\mathcal{L}(p_i,z_j|D)=\prod_{\substack{j=1\\\sum_k c_{jk} > 0}}^N\frac{\prod_{k=1}^{m_j}\left[c_{jk}p_{r_{jk}}z_j+(1-c_{jk})(1-p_{r_{jk}}z_j)\right]}{1-\prod_{k=1}^{m_j}(1-p_{r_{jk}}z_j)}\,.
\end{equation}

From now on we will assume by default that our dataset contains only query-document pairs that received at least one click and will omit the subscript $\sum_k c_{jk} > 0$.

Our last step will be to simplify the likelihood function (\ref{likelihood-1}). Typically the click probabilities $p_iz_j$ are not very large (i.e. not close to $1$). This is the probability that the query-document pair $j$ will get a click when displayed at rank $i$. To simplify the likelihood for each query-document pair we will only keep terms linear in $p_iz_j$ and drop higher order terms like $p_{i_1}z_{j_1}p_{i_2}z_{j_2}$. We have verified this simplifying assumption for our data in Section \ref{sec-ebay}. In general, we expect this assumption to be valid for most search engines. It is certainly a valid assumption for lower ranks since click through rates are typically much smaller for lower ranks. Since we are dropping product terms the largest ones would be between ranks $1$ and $2$. For most search engines the click through rates at rank 2 are around 10\% or below, which we believe is small enough to be able to safely ignore the product terms mentioned above (they would be at least $10$ times smaller than linear terms). We empirically show using simulations in Appendix \ref{section-sim} that this assumption works very well for data similar to eBay data. If for other search engines the click through rates are much larger for topmost ranks we suggest keeping only those query-document pairs that appeared at least once at a lower enough rank. Also, using the methodology of simulations from Appendix \ref{section-sim} one can verify how well this assumption works for their particular data.

Under the simplifying assumption we get for the denominator in (\ref{likelihood-1}):
\begin{equation}\label{simplify-denom}
	1-\prod_{k=1}^{m_j}(1-p_{r_{jk}}z_j)\simeq1-\left(1-\sum_{k=1}^{m_j}p_{r_{jk}}z_j\right)=z_j\sum_{k=1}^{m_j}p_{r_{jk}}\,.
\end{equation}

Let us now simplify the numerator of (\ref{likelihood-1}). Firstly, since the click probabilities are not large and each query-document pair appears only a few times we can assume there is only one click per query-document pair\footnote{This is true for our data as discussed in Section \ref{sec-ebay}. For the cases when most query-document pairs receive multiple clicks we suggest using a different method, such as computing the ratios of propensities by computing the ratios of numbers of clicks as discussed in Appendix \ref{sec-ratio}.}. We can assume $c_{jl_j}=1$ and $c_{jk}=0$ for $k\neq l_j$. The numerator then simplifies to
\begin{align}\label{simplify-numer}
	\begin{split}
		\prod_{k=1}^{m_j}\left[c_{jk}p_{r_{jk}}z_j+(1-c_{jk})(1-p_{r_{jk}}z_j)\right]&=p_{r_{jl_j}}z_j\prod_{\substack{k=1\\k\neq l_j}}^{m_j}(1-p_{r_{jk}}z_j)\\
											     &\simeq p_{r_{jl_j}}z_j\,.
	\end{split}
\end{align}

Using (\ref{simplify-denom}) and (\ref{simplify-numer}) the likelihood function (\ref{likelihood-1}) simplifies to
\begin{equation}\label{likelihood-final}
	\mathcal{L}(p_i,z_j|D)=\prod_{j=1}^N\frac{p_{r_{jl_j}}z_j}{z_j\sum_{k=1}^{m_j}p_{r_{jk}}}=\prod_{j=1}^N\frac{p_{r_{jl_j}}}{\sum_{k=1}^{m_j}p_{r_{jk}}}\,.
\end{equation}

In the last step $z_j$ cancels out from the numerator and the denominator. Our assumption of small click probabilities, together with keeping only query-document pairs that received at least one click allowed us to simplify the likelihood function to be only a function of propensities. Now we can simply maximize the likelihood (\ref{likelihood-final}) to estimate the propensities.

Eq. (\ref{likelihood-final}) makes it clear why we need to include the requirement that each query-document pair should appear more than once at different ranks. If we have a query-document pair that appeared only once (or multiple times but always at the same rank) then the numerator and the denominator would cancel each other out in (\ref{likelihood-final}). For that reason we will keep only query-document pairs that appeared at two different ranks at least.

It is numerically better to maximize the log-likelihood, which takes the form:
\begin{equation}\label{log-like-appendix}
	\log\mathcal{L}(p_i|D)=\sum_{j=1}^N\left(\log(p_{r_{jl_j}})-\log\sum_{k=1}^{m_j}p_{r_{jk}}\right)\,.
\end{equation}

\section{Propensity Ratio Estimation}\label{sec-ratio}

Here we consider the case when for two fixed ranks $i$ and $j$ a large number of query-document pairs appear at both of these ranks. As mentioned in Section \ref{section-method} one can simply compute the number of clicks from those query-document pairs for each rank and take the ratio of those numbers to estimate the ratio $p_i/p_j$. We prove that statement below.

Let us assume that query-document pairs $\{x_k, k = 1,\dots,K\}$ appeared for both ranks $i$ and $j$. We will first assume that these query-document pairs appeared exactly once for each rank. We will later relax that assumption. The probability of a click at rank $i$ for query-document pair $x_k$ is $p_iz_k$ where $z_k$ is the ``relevance'' for query-document pair $x_k$ (i.e. the probability of a click under the assumption that it was observed). So the expected number of clicks $N_i$ for rank $i$ for all query-document pairs will be:
\[
	\mathbb{E}[N_i]=\sum_{k=1}^Kp_iz_k=p_i\sum_{k=1}^Kz_k\,.
\]

Taking the ratio of the above for $i$ and $j$\footnote{Note that here we are taking the ratio of expected values rather than the expected value of the ratio. While in general they are not the same, to the first order approximation they are the same. We have empirically verified through simulations that taking the ratio of the observed values $N_i/N_j$ asymptotically approaches the ratio $p_i/p_j$ as the total number of query-document pairs is increased.}:
\[
	\frac{\mathbb{E}[N_i]}{\mathbb{E}[N_j]}=\frac{p_i\sum_{k=1}^Kz_k}{p_j\sum_{k=1}^Kz_k}=\frac{p_i}{p_j}\,.
\]

Replacing the expected values of $N_i$ and $N_j$ above by their actual observed values we can get an estimator for $p_i/p_j$.

So far we have assumed that each query-document pair appears exactly once at each rank. The above proof can be easily extended to the case when the same query-document pair appears multiple times at each rank. In this case instead of counting clicks directly we will count clicks per impressions, i.e. for each query-document pair we will count the total number of clicks and divide by the total number of impressions. The expected value for clicks per impressions is still the click probability $p_iz_k$. Then we will take the sum for clicks per impressions for each rank and take the ratio. The derivation will remain exactly the same. The only difference is that $N_i$ and $N_j$ will now denote the sum of clicks per impressions instead of the total number of clicks.

\section{Results on Simulations}\label{section-sim}

\begin{figure}

\begin{center}
	\includegraphics[width=8cm]{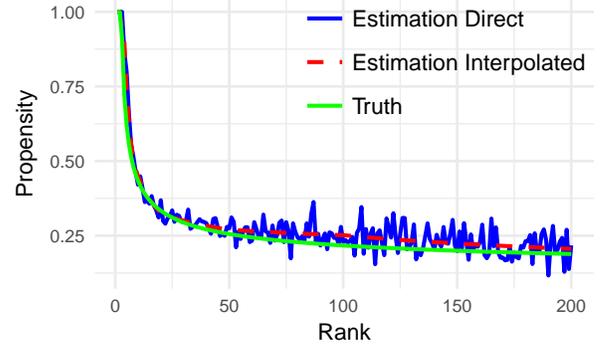}
\end{center} 

\setlength{\belowcaptionskip}{-12pt}
\caption{Propensity estimated from simulated data. The green solid curve shows the ``true'' propensity (\ref{prop-true}). The blue solid curve is the estimated propensity using the direct estimation method. The red dashed curve is the estimation using interpolation.}\label{fig-prop-simulation}
\end{figure}

In this Appendix we use simulated data to verify that the method of estimating propensities developed in Section \ref{section-method} works well. For our simulations we choose the following propensity function as truth:
\begin{equation}\label{prop-true}
	p_i^{\mathrm{sim}}=\min\left(\frac{1}{\log{i}},1\right)
\end{equation}
which assigns propensity of $1$ for ranks $1$ and $2$, and then decreases as the inverse of the log of the rank.

Other than choosing our own version of propensities we simulate the data to be as similar to the eBay dataset as possible. We generate a large number of query-document pairs and randomly choose a mean rank $rank_{mean}$ for each query-document pair uniformly between 1 and 500. We randomly generate a click probability $z$ for that query-document pair depending on the mean rank $rank_{mean}$. We choose the distribution from which the click probabilities are drawn such that the click through rates at each rank match closely with the click through rates for real data, taking into account the ``true'' propensities (\ref{prop-true}). We then generate two different ranks drawn from $\mathcal{N}(rank_{mean}, (rank_{mean} / 5)^2)$. For each rank $i$ we compute the probability of a click as $zp_i^{\mathrm{sim}}$. Then we keep only those query-document pairs which appeared at two different ranks and got at least one click, in agreement with our method used for real eBay data. Finally, we keep about 40,000 query-document pairs so that the simulated data is similar to the eBay web search data in size. This becomes the simulated data.

The estimated propensities on the simulated dataset are shown in Fig.~\ref{fig-prop-simulation}. The green solid curve shows the true propensity (\ref{prop-true}), the blue solid curve shows the estimated propensity using the direct estimation method, and the red dashed curve is the estimated propensity using interpolation. As we can see, the estimations closely match with the truth. Furthermore, we can see that the interpolation method gives a better result by reducing the noise in the estimate. These results show that the propensity estimation method developed in this paper works well.

\bibliographystyle{ACM-Reference-Format}
\bibliography{bibliography}

\end{document}